\documentstyle[preprint,aps,psfig]{revtex}

\def\Tr{{\rm Tr}}
\def\lsim{\mathrel{\mathpalette\oversim<}}

\def\oversim#1#2{\lower0.2ex\vbox{\baselineskip0pt\lineskip0pt
  \lineskiplimit0pt\ialign{$#1\hfil##\hfil$\crcr#2\crcr\sim\crcr}}}
\def\ns{{\ooalign{\hfil/\hfil\crcr$\nabla$}}}

\begin{document}

\draft
\preprint{HUPD-9832}
\title{A Model of Curvature-Induced Phase Transitions\\
       in Inflationary Universe}
\author{J. Hashida, S. Mukaigawa, T. Muta, K. Ohkura and K. Yamamoto}
\address{Department of Physics, Hiroshima University\\
         Higashi-Hiroshima, Hiroshima 739-8526}

\date{\today}
\maketitle
\begin{abstract}

Chiral phase transitions driven by space-time curvature effects
are investigated in de Sitter space in the supersymmetric
Nambu-Jona-Lasinio model with soft supersymmetry breaking.
The model is considered to be suitable for the analysis of possible
phase transitions in inflationary universe.
It is found that a restoration of the broken chiral symmetry takes place
in two patterns for increasing curvature : the first order and second order
phase transition respectively depending on initial settings of
the four-body interaction parameter and the soft supersymmetry
breaking parameter. The critical curves expressing the phase boundaries
in these parameters are obtained. Cosmological implications of the result
are discussed in connection with bubble formations and the creation of
cosmic strings during the inflationary era.

\end{abstract}


\narrowtext

In the scenario of the early universe the Higgs mechanism is one of the
possible candidates in explaining the onset of the inflation era.
At the beginning of the inflation era
it is assumed that the grand unified theory phase is broken down to the
quantum chromodynamics and electroweak theory phase through the Higgs
mechanism. In this connection it is interesting to note that the Higgs fields
may be composed of some fundamental fermions as in the technicolor model and
to see the consequence of this idea in the scenario of the inflation.
On the other hand the supersymmetry is considered to be a vital nature possessed
by the fundamental unified theory and hence the incorporation of the
supersymmetry in composite Higgs models seems to be of principal importance.
Under these circumstances it is natural for us to consider a supersymmetric
composite Higgs model in the early stage of the universe and to see whether
any remarkable effects are drawn during the inflation era.
For simplicity we adopt the Nambu-Jona-Lasinio (NJL) model \cite{Nam} as a
prototype of the composite Higgs model in the present communication.

At the inflation era the quantum effect of the gravitation is of minor
importance while the external gravitational field is non-negligible. Hence we
are naturally led to the supersymmetric NJL model in curved space.
Dealing with the composite Higgs fields is essentially nonperturbative
and does not accept approximate treatments. Accordingly we try to solve the
problem rigorously working in a specific space-time, the de Sitter space,
which possesses a maximal symmetry. The de Sitter space is suitable for
describing the inflationary universe. As a nonperturbative method we rely on
the $1/N$ expansion technique.

The four-fermion interaction model (which is the basis of the NJL model)
in de Sitter space has been discussed by several authors
\cite{Ito,Bu1,Ina,Eli} and is found to reveal the restoration of the
broken chiral symmetry for increasing curvature as a second order phase
transition. The supersymmetric version of the NJL model in curved space
was considered by I. L. Buchbinder, T. Inagaki and S. D. Odintsov \cite{Bu2}
in the weak curvature limit. They found that the chiral symmetry is broken
as the curvature increases. Their result is in contrast with the result in
the nonsupersymmetric NJL model. On the other hand the supersymmetric NJL
model in the flat space-time has been investigated by several authors \cite{Lov}
in the context of dynamical chiral symmetry breaking.

We start with the Lagrangian for the supersymmetric Nambu-Jona-Lasinio model
in de Sitter space expressed in terms of component fields of
the superfields \cite{Bu2}, 

\begin{eqnarray}
 {\cal L} &=& -\nabla^{\mu}\phi^{\dag} \nabla_{\mu}\phi -\rho^2 \phi^{\dag}\phi
 -\nabla^{\mu}\phi^{c \dag} \nabla_{\mu}\phi^{c} -\rho^2 \phi^{c \dag}\phi^{c} 
 \nonumber
 \\
 &&-\bar{\psi}(i\ns -\sigma + i\gamma_{5}\pi)\psi -\frac{N}{2\lambda}\rho^2, 
\label{L}
\end{eqnarray}
where we have kept only terms relevant to the leading order in the $1/N$ 
expansion. In Eq.~(\ref{L}) $\phi$ and $\psi$ refer to the scalar and 
spinor component fields of the superfield respectively, $N$ is the 
number of components of these fields, $\nabla_{\mu}$ the covariant 
derivative, $\lambda$ the four-fermion coupling constant
and $\rho$ the auxiliary scalar field with $\rho^2\equiv \sigma^2+\pi^2 $.
We introduce an additional term $\delta{\cal L}$ composed of non-minimal
gravitational terms and a soft supersymmetry breaking term to
the above Lagrangian (\ref{L})
\begin{equation}
\delta{\cal L} = -\xi_1 R \phi^{\dag}\phi - \xi_2 R \phi^{c \dag}\phi^{c}
-\Delta^2 (\phi^{\dag}\phi + \phi^{c \dag}\phi^{c}),
\label{dL}
\end{equation}
where $R$ is the space-time curvature and $\xi_1$, $\xi_2$ and $\Delta$ are
coupling parameters. We assume that
the term (\ref{dL}) existed already when the inflation era started.

The effective potential for the auxiliary field $\rho$ is calculated in the
leading order of the $1/N$ expansion such that \cite{Bu2}

\begin{eqnarray}
 &&V(\rho) = \frac{\rho^2}{2\lambda} 
 +i \int^{\rho}_{0} ds \Tr S(x, x; s)
 \nonumber
 \\
 &&\hspace{1.8cm}+2i \int^{\rho^2}_{0} dt G(x, x; t) \label{V}
\end{eqnarray}
where $S(x,y;s)$ and $G(x,y;t)$ represent the fermion and boson propagator
in the coordinate space with mass $s$ and $\sqrt{t}$ respectively.
The effective potential (\ref{V}) is obtained by taking the short-distance
limit $y\to x$ once the full expressions of these propagators are found
in de Sitter space.

The boson propagator in de Sitter space is well-known
\cite{BirrellDavies,BunchDavies,Allen,AllenJ,Tagirov} and is given
for arbitrary dimension $D$ by
\begin{equation}
  G(x,y;t)=-i{r^{2-D} \over (4\pi)^{D/2}}
  {\Gamma(a_+)\Gamma(a_-)\over\Gamma(D/2)}
  {}_2F_1(a_+, a_-, D/2; 1-z), \label{B}
\end{equation}
with $a_{\pm}=(D-1\pm \sqrt{(D-1)^2-4tr^2})/2$.
Here $z=\sigma^2/(2r^2)$ with $\sigma$ the geodesic distance and $r$
the radius of de Sitter space which is related to the conventional
Hubble parameter such that $r=1/H$. 
The scalar propagator for Lagrangian
${\cal L}+\delta{\cal L}$ is obtained simply by replacing $t$
by $t+\xi R +\Delta^2$ with $\xi=\xi_1=\xi_2$ and $R=D(D-1)/r^2$ where
we have taken $\xi_1=\xi_2$ for simplicity.
The fermion propagator is given by \cite{Mukaigawa,Campo}
\begin{equation}
  S(x,y;s)=\bigl(A(x,y;s)+B(x,y;s)\sigma_{;\mu}\gamma^\mu\bigr)U, \label{F}
\end{equation}
where
\begin{eqnarray}
 A(x,y;s)&=&i{sr^{2-D} \over (4\pi)^{D/2}}
  {\Gamma(a)\Gamma(a^*)\over\Gamma(D/2+1)}
\nonumber
\\
  &&\times\sqrt{1-z}{}_2F_1(a, a^*, D/2+1; 1-z), \label{FA}
\end{eqnarray}
and $a=D/2+isr$ and $U$ is the matrix composed of the Dirac matrices.
We do not present the explicit expression of the invariant function
 $B(x, y; s)$ although it is known analytically. The reason is that we do
not need the explicit expression of this function since the second term
 on the right hand side of Eq.~(\ref{F}) disappears for vanishing distance.
We find for small geodesic distance $z\sim 0$ that
\begin{eqnarray}
 \Tr S(x,x;s)&=&\lim_{z\rightarrow0}A(x,y;s)\Tr U
 \nonumber
 \\
 &=&\Tr[1]\lim_{z\rightarrow0}A(x,y;s),
 \label{S0}
\end{eqnarray}
with the normalization $\Tr U=\Tr[1]$.
Equipped with these propagators for bosons and fermions we are now ready to
calculate the effective potential (\ref{V}) in an exact form. 

In order to explain our idea in a transparent way we mainly work
in $3$ space-time dimensions and give a brief comment on the full extension
to $4$ dimensions.
We use a well-known formula which is found in any mathematical table
(e.g., \cite{Magnus}) to rewrite the boson propagator and find
for small geodesic distance $z\sim 0$,
\begin{equation}
  G(x,y;t)={-i\over 8\pi rz^{1/2}}+{i\over 4\pi r} \nu {\rm coth} \pi\nu
  +{\cal O}(z^{1/2})~, \label{G03}
\end{equation}
where we have defined $\nu=\sqrt{tr^2+6\zeta-1}~$ and 
$\zeta=\xi+\Delta^2r^2/6$.
Note here that we are calculating the boson propagator
for Lagrangian ${\cal L}+\delta {\cal L}$.
The fermion propagator at vanishing distance in $3$ dimensions is obtained
in a similar way. Using the relation (\ref{S0}) we have
\begin{eqnarray}
  \Tr S(x,x;s)&=&\Tr[1]\biggl[ {is\over 8\pi rz^{1/2}}
  -{is\over 4\pi r}\biggl({1\over4}+s^2 r^2\biggr) 
  {{\rm tanh} \pi s r\over s r}
\nonumber
\\
  &&\hspace{3cm}  +{\cal O}(z^{1/2})~\biggr], \label{S03}
\end{eqnarray}
Substituting propagators (\ref{G03}) and (\ref{S03}) into
effective potential (\ref{V}) we find
\begin{eqnarray}
  &&V(\rho)={\rho^2\over 2\lambda}
  +{1\over \pi r}\int_0^{\rho} ds s  \biggl({1\over 4}+s^2 r^2\biggr)
  {\tanh \pi s r\over sr}
\nonumber
\\
  &&\hspace{1.8cm}
  -{1\over 2\pi r}\int_0^{\rho^2} dt \nu {\rm coth} \pi \nu, \label{V3}
\end{eqnarray}
where we have set $\Tr[1]=4$ (We adopt the reducible representation of the
Clifford algebra of Dirac matrices to afford the existence of $\gamma_5$).
It is important to note here that the divergence present in both expressions
(\ref{G03}) and (\ref{S03}) cancel out in the effective potential (\ref{V3}).
The origin of this cancellation may be traced back to the supersymmetry of
our model.

The gap equation is given by
\begin{eqnarray}
 &&V'(\rho^2)={1\over 2\pi r}
 \biggl[{\pi r\over \lambda}
 +\biggl({1\over4}+\rho^2r^2\biggr){\tanh\pi \rho r \over \rho r}
\nonumber
\\
 &&\hspace{0.5cm}-\sqrt{\rho^2 r^2+6\zeta-1}
  {\rm coth}{(\pi \sqrt{\rho^2 r^2+6\zeta-1})}
 \biggr]=0~, \label{Gap3}
\end{eqnarray}
where by $V'(\rho^2)$ we mean the differentiation of the effective
potential with respect to $\rho^2$.
Just by observing the left hand side of the gap equation (\ref{Gap3}) we
find that it is a function only of $\rho r$ with parameters 
$\pi r/\lambda$ and $\zeta$.
Thus the solution $\rho r$ of this gap equation is completely
specified by two parameters $\pi r/\lambda$ and $\zeta$
respectively. This fact suggests that
the phase diagram of the model will be given on a plane specified by
these two parameters. 

The direct observation of Eq.~(\ref{Gap3}) shows that it has at most
two solutions for $\rho r$ and so the shape of the effective potential
(\ref{V3}) is of three types: the symmetric type S and type I and II
which are shown in Fig.~1.
The phase diagram of the model is given by the numerical analysis of
the effective potential (\ref{V3}) and the gap equation (\ref{Gap3}),
and is given in Fig.~2.
The region above the dashed line and the right half of the solid line
in Fig.~2 represents a broken phase with the effective potential of
the shape II given in Fig.~1. 
The small region between the solid line
and the dashed line in Fig.~2 corresponds to a broken phase with 
the effective potential of the shape I in Fig.~1.
The region below the whole solid line represents a symmetric phase
where the shape of the effective potential is of the single-well type S.

The boundaries which separate the above three phases are
determined as follows: The dashed line and the right half of the solid
line constitute the boundary of the region characterized by the potential
of the shape II. The boundary is determined by the condition
\begin{equation}
  V'(0)=0.
\end{equation}
The above equation reduces to
\begin{equation}
  {\pi r\over \lambda}+{\pi\over 4}-\sqrt{6\zeta-1}\coth\pi
\sqrt{6\zeta-1}=0. \label{d1V3}
\end{equation}
The condition for the line separating the phases of type I and type S
is given by $V(\rho_*)=0$
where $\rho_*={\rm max}\{\rho_1,\rho_2\}$ with $\rho_1$ and $\rho_2$
two solutions of the gap equation (\ref{Gap3}).

The branching point C in Fig.~2 is of special interest. It is a critical
point which divides the broken phase into type I and type II. At the
branching point C the following conditions are found to be satisfied
simultaneously:
\begin{equation}
     V'(0)=0, \ \ \ V''(0)=0. \label{d2V3}
\end{equation}
The condition (\ref{d2V3}) is explicitly given by Eq.~(\ref{d1V3}) and
\begin{equation}
 1-\frac{\pi^2}{12}+\frac{1}{2\sinh^2\pi\sqrt{6\zeta-1}}
 -\frac{\coth\pi\sqrt{6\zeta-1}}{2\pi\sqrt{6\zeta-1}}=0. 
 \label{Crit3}
\end{equation}
From Eq.~(\ref{Crit3}) we find that $\zeta=0.290138 (\equiv \zeta_*)$ 
at the branching point C. By substituting this value for $\zeta$ 
in Eq.~(\ref{d1V3}) we obtain $\pi r/\lambda=0.083063 (\equiv \eta_*)$
at the branching point C.

Now let us discuss the time-evolution of the chiral structure of the model
assuming that the inflationary era is well described by the
effective potential in de Sitter space. 
It is natural to assume that the curvature slowly decreases ($r$ 
increases) as the universe evolves. First let us consider the 
case where the soft supersymmetry breaking term is not included, 
i.e., $\Delta=0$. 
In this case it is easily seen in Fig.~2 that we move from left to 
right by increasing radius $r$ with $\zeta(=\xi)$ and $\lambda$ fixed. 
By direct observation of the gap equation one can easily show that,
if the parameter $\xi$ is kept below $\xi=1/4$ which is the value of
$\xi$ corresponding to $\pi r/\lambda=0$, the effective 
potential stays in the type S as $r$ increases and
 hence the chiral symmetry is preserved (there is no
phase transition). If the parameter $\xi$ is kept in the region
$1/4<\xi<\zeta_*$, the effective potential changes its shape from 
type I to type S as $r$ increases. Hence the broken chiral symmetry is
restored as $r$ increases through the first order phase transition.
If the parameter $\xi$ is kept above $\zeta_*$,
the effective potential changes its shape from type II to type S
and so the transition is of the second order. Thus in the case
with $\Delta=0$ the chiral symmetry restoration occurs as the universe
evolves. This situation reflects the fact that the non-minimal
gravitational coupling terms, which break the supersymmetry
that protects the chiral symmetry for any value of $\lambda$
and $r$, are effective only at large curvature.
It is also interesting to examine the pattern of the phase transitions
when the parameter $\xi$ is changed with $r$ and $\lambda$ fixed.
In this case we easily conclude that the chiral symmetry is broken
by increasing $\xi$ through the first order phase transition
for $\pi r/\lambda<\eta_*$ while the transition is of the
second order for $\pi r/\lambda>\eta_*$.

We next focus our attention on the role of the soft supersymmetry
breaking terms. Here we assume the conformal gravitational coupling
in 3 dimension, i.e., $\xi=1/8$, for simplicity. 
This assumption is not essential for discussions in the following
as long as $\xi<1/4$. 
In Fig.~3 we show trajcetories on 
the phase diagram as the radius $r$ increases with $\Delta$ and $\lambda$ 
fixed (See dot-dashed lines (a)-(c) in Fig.~3). The trajectories are
specified by the equation $\zeta=1/8+(\Delta\lambda/\pi)^2\eta/6$ with
$\eta=\pi r/\lambda$ for each value of $\Delta\lambda/\pi$.
Increasing the radius $r$ along the curve (a) one experiences a first
order phase transition and along the curve (c) a second order one.
The critical case is shown by the curve (b). These curves show that the 
chiral symmetry breaking can occur in the model with the soft supersymmetry 
breaking terms when the radius $r$ increases or when the curvature of the 
universe decreases during the inflation. By observing the behaviors of
these curves we find that the types of the phase transitions are
classified as follows: If 
$\Delta\lambda/\pi>\alpha(\equiv 29.5\sqrt{\zeta_*-1/8}=12.0)$ is satisfied,
the phase transition is of the first order.
The case $1<\Delta\lambda/\pi<\alpha$ corresponds to the second order
phase transition, and the phase transition does not occur
if $\Delta\lambda/\pi<1$ is satisfied. 

In summary we have investigated the supersymmetric NJL 
model with the non-minimal gravitational coupling terms and 
soft supersymmetry breaking terms in de Sitter space. 
The phase structure is completely clarified in the model of 3 dimensions. 
We have found that both the non-minimal gravitational terms and the 
soft supersymmetry breaking terms lead to the phase transition phenomena
as radius $r$ increases or the curvature of the universe
decreases. These two terms work in different ways.
The soft supersymmetry breaking terms lead to the chiral symmetry 
breaking as the curvature decreases while the non-minimal
gravitational terms lead to the restoration of the chiral symmetry. 
The extension of our work to the 4 dimensional case is straightforward
although we have to rely on the full numerical estimates. The result of
the analysis will be presented elsewhere.

The cosmological consequences of the phase transitions found here
seems to be quite interesting.
A production of cosmic strings is expected because the $U(1)$-symmetry
is broken in the model.
The formation of topological defects in the inflationary universe
has been studied in many references (e.g., \cite{YokoyamaNagasawa}).
It is well-known that
the cosmic strings are constrained observationally as 
$G\mu\lsim {\cal O}(10^{-5})$ where $G$ is the gravitational 
constant and $\mu$ is the line density of the cosmic string \cite{Kolb}.
This condition sets a constraint on our model and it is an interesting
problem to investigate whether the condition is satisfied in the model.
In order to apply our model to
a more realistic inflationary scenario we need to deal with
the model in 4 dimensions. At the moment we have found 
that the phase structure in 4 dimensions is very similar to
that in 3 dimensions. The expression of the effective 
potential in 4 dimensions is rather cumbersome and the detailed
analysis is still in progress.

Another comment is directed to the possibility of creation 
of an open universe in our model. 
Inflation models for an open universe have been investigated
by using a bounce solution for a bubble formation \cite{Open}.
The trajectory (a) in Fig.~3 indicates that bubble nucleations 
may occur in the region I. It is worthy to investigate whether 
our model can provide a successful model for an open universe or not.

The authors would like to thank Roberto Camporesi, Atsushi Higuchi,
Tomohiro Inagaki and Misao Sasaki for enlightening discussions and
useful correspondences.
Two of the authors (T. M. and K. Y.) are indebted to Monbusho Fund
(Grant-in-Aid for Scientific Research (C) from the Ministry of Education,
Science and Culture with contract numbers 08640377 and
09740203 respectively) for financial supports.


\begin{figure}
\begin{center}
    \leavevmode\psfig{file=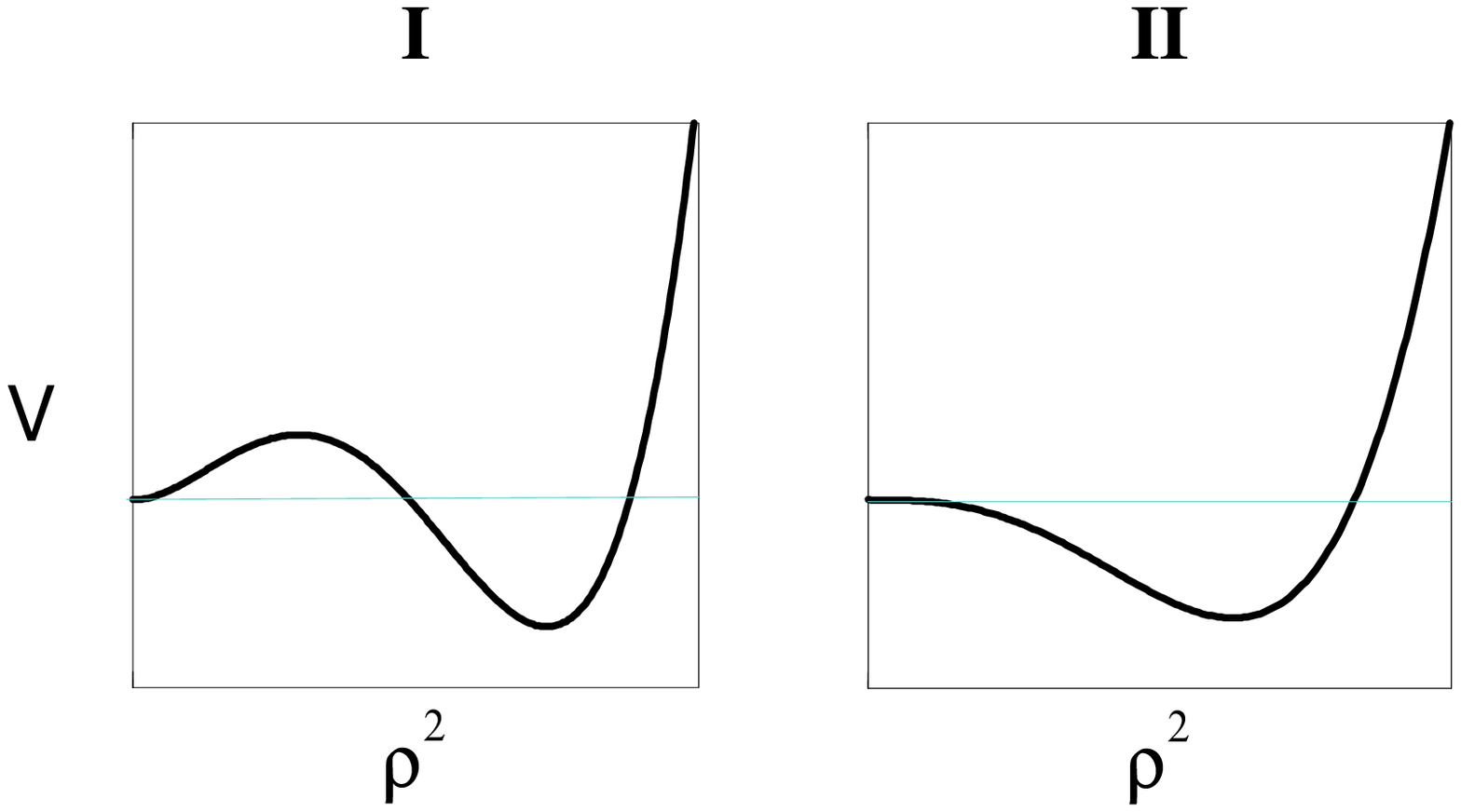,width=8cm}
\end{center}
\caption{Typical behaviors I and II of the effective potential.
\label{fig:potential}}
\end{figure}
\begin{figure}
\begin{center}
    \leavevmode\psfig{file=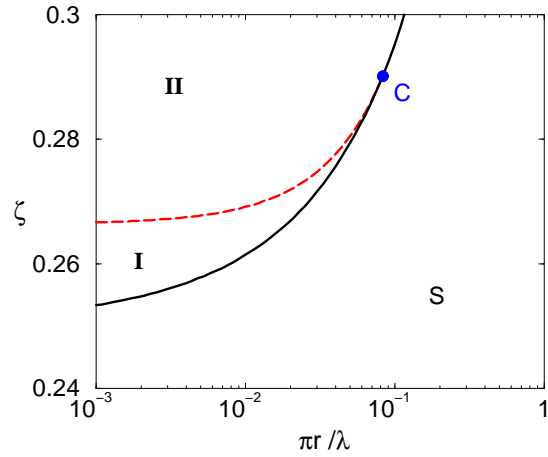,width=8cm}
\end{center}
\caption{The phase diagram on the parameter plane in $\zeta$
          and $\pi r/\lambda$. The point denoted by C is the branching point
          where $(\pi r/\lambda, \zeta)=(\eta_*, \zeta_*)$ with
          $\eta_*=0.083063$ and $\zeta_*=0.290138$.
         \label{fig:phase}}
\end{figure}
\begin{figure}
\begin{center}
    \leavevmode\psfig{file=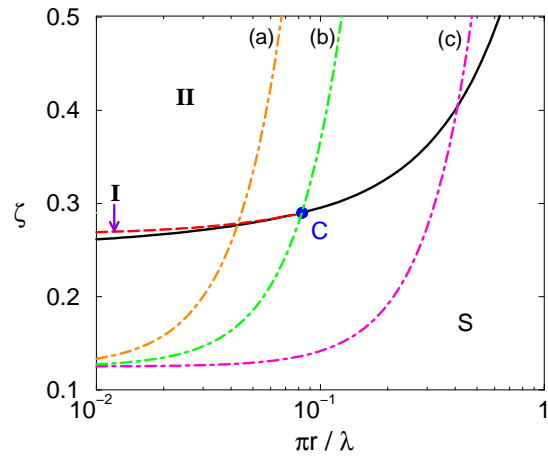,width=8cm}
\end{center}
\caption{Typical trajectories for varying
 $r$ with $\Delta$ and $\lambda$ fixed,
(a) $\Delta\lambda/\pi=22.3$; (b)$\Delta\lambda/\pi=12.0$;
(c) $\Delta\lambda/\pi=3.2$.
Here $\xi$ is fixed so that $\xi=1/8$.
\label{fig:traj}}
\end{figure}
\end{document}